\documentclass[final,3p,times]{elsarticle}
\usepackage{amsmath,amsfonts,amssymb,array}
\usepackage{enumerate}
\usepackage{float}
\usepackage{graphicx}
\usepackage{multirow}
\usepackage{color}
\usepackage[colorlinks = true,linkcolor = blue]{hyperref} 
\newcommand{\CPV}{\text{CP}\!\!\!\!\!\!\!\raisebox{0pt}{\small$\diagup$}}
\newcommand{\equaref}[1]{Eq.~(\ref{#1})}
\newcommand{\equasref}[2]{Eqs.~(\ref{#1})~and~(\ref{#2})}

\newcommand{\figref}[1]{Fig.~\ref{#1}}

\usepackage{lipsum}  
\begin{document}
\leftline{}
\rightline{IPPP/16/83}

\setcounter{footnote}{0}
\renewcommand{\thefootnote}{\fnsymbol{footnote}}

\title{Baryogenesis via leptonic CP-violating phase transition}

\author{
Silvia Pascoli\,\footnote{Email: silvia.pascoli@durham.ac.uk}$^{a}$,~~
Jessica Turner\,\footnote{Email: jturner@fnal.gov}$^{a,b}$~~and~~
Ye-Ling Zhou\,\footnote{Email: ye-ling.zhou@durham.ac.uk}$^{a}$
\vspace*{3mm}
}

\address{
$^a$\,Institute for Particle Physics Phenomenology, Department of Physics, Durham University, Durham DH1 3LE, U.K.
\\[1mm]
$^b$\,Theoretical Physics Department, Fermi National Accelerator Laboratory,
P.O. Box 500, Batavia, IL 60510, USA
}

\date{\today}
%*****************************************************************************

\begin{abstract}
We propose a new mechanism to generate a  lepton asymmetry based on the vacuum CP-violating phase transition (CPPT). This approach differs from 
 classical thermal leptogenesis as a specific seesaw model, and its UV completion, need not be specified.  The lepton asymmetry is generated via the dynamically 
  realised coupling of the Weinberg operator during the phase transition. This mechanism provides a
  connection with low-energy neutrino observables. 
  %strong connections with low-energy neutrino experiments. 
\\[2.5mm]
\end{abstract}

\maketitle

\renewcommand{\thefootnote}{\arabic{footnote}}
\setcounter{footnote}{0}
%%%%%%%%%%%%%%%%%%%%%%%%%%%%%%%%%%%%%%%%%%%%%%%%%%%%%%%%%%%%%%%%%%%%%%%%%%%%%%

\section{Introduction}
The origin of the matter-antimatter asymmetry is one of the most important mysteries of our Universe.   
One popular mechanism to explain this asymmetry  is baryogenesis via leptogenesis. The classic examples of high-scale leptogenesis \cite{Fukugita:1986hr} introduce  right-handed neutrinos, $N$, which are Majorana in nature and therefore break lepton number. The decay of the right-handed neutrinos provides a departure from equilibrium and their couplings to  leptonic doublets, 
$\ell$, violate CP. A lepton asymmetry, produced from the preferential decays of $N$,  is subsequently converted to a baryon asymmetry by $B-L$ conserving sphaleron processes \cite{Khlebnikov:1988sr}. In addition to fulfilling Sakharov's criterion \cite{Sakharov:1967dj}, this scenario of leptogenesis provides a natural explanation of small neutrino masses via the seesaw mechanism. 

The origin and energy scale of CP violation is still unknown and remains a widely studied theoretical issue. There is a rich programme of neutrino experiments such as LBNF/DUNE \cite{DUNE} and T2HK \cite{T2HK} that aim to measure leptonic CP violation. In conjunction, these experiments will investigate the correlations between leptonic observables. The interrelation between mixing angles and phases will play a crucial role in determining the fine structure of leptonic mixing. The observed pattern may be the result of an underlying flavour symmetry which could be continuous, $U(1)$ \cite{Froggatt:1978nt}, $SU(3)$, $SO(3)$ \cite{Alonso:2013nca},  or a non-Abelian, discrete symmetry such as $A_4$, $S_4$ \cite{King:2014nza}, {\it et al}.  In these models, SM-singlet scalars (flavons) acquire vacuum expectation values that lead to the breaking of the flavour symmetry and results in the observed mixing structure and CP violation. The source of CP violation can arise  \emph{spontaneously} or \emph{explicitly}. Spontaneous CP violation  refers to the scenario in which CP conservation is imposed on the Lagrangian but is spontaneously broken by the vacuum \cite{Branco:2011zb, deMedeirosVarzielas:2011zw} whilst explicit CP violation results from complex Yukawa couplings. Unlike conventional high-scale leptogenesis, where CP is explicitly violated above the seesaw scale, in this work we investigate the possibility CP violation occurs below such a scale.  

In this paper, we propose a new mechanism for generating a lepton asymmetry based on a  CP-violating phase transition (CPPT). This mechanism allows  a connection between the baryon asymmetry, neutrino oscillation experiments and leptonic flavour mixing. CPPT differs from conventional scenarios of high-scale leptogenesis in several key aspects: we apply an effective field theory approach, which does not constrain the study to a particular model of neutrino mass generation and consequently CP violation occurs below this energy scale.
 We simply  assume that neutrino masses are generated by the Weinberg operator, the coefficients of which are dynamically realised during CPPT. 
The lepton asymmetry is produced via the interference of Weinberg operators at different times. To perform the calculation we  utilise the closed-time-path (CTP) formalism \cite{Schwinger:1960qe, Keldysh:1964ud}. We focus on the generation of initial asymmetry at the constant temperature, $T$, and defer a  more complete calculation of the final lepton asymmetry, accounting for evolution, to future work \cite{future}. 

%%%%%%%%%%%%%%%%%%%%%%%%%%%%%%%%%%%%%%%%%%%%%%%%%%%%%%%%%%%%%%%%%%%%%%%%%%%%

\section{The Mechanism }
As previously mentioned, throughout this work, we assume neutrinos acquire their masses via the dimension-5 Weinberg operator
\begin{eqnarray}\label{eq:Weinberg}
\mathcal{L}_{W}=\frac{\lambda_{\alpha\beta}}{\Lambda}\ell_{\alpha L}H C\ell_{\beta L} H + \text{h.c.} \,, 
\end{eqnarray}
where $\lambda_{\alpha\beta}=\lambda_{\beta\alpha}$ and $C$ is the charge conjugation matrix. 
For the purpose of generating a lepton asymmetry, the full UV completion of this operator need not be specified. The coupling of the Weinberg operator, $\lambda_{\alpha\beta}$, can be associated to a SM-singlet scalar field (or a linear combination of scalar fields) whose vacuum expectation value (VEV) corresponds to physical  leptonic masses and mixing. This dynamically generated  coupling may be realised as  $\lambda_{\alpha\beta}=\lambda^{0}_{\alpha\beta}+\lambda^{1}_{\alpha\beta}\langle\phi\rangle/v_{\phi}$,  where $\lambda^{0}_{\alpha\beta}$ are initial values of $\lambda_{\alpha\beta}$ before the flavon $\phi$ undergoes a phase transition. 

In the Early Universe, the ensemble expectation value (EEV) of $\phi$ is dependent upon the finite temperature  scalar potential. 
A phase transition occurs when the minima of this potential becomes metastable. As a consequence, the minimum changes  to a non-zero and stable value, $\langle \phi \rangle$. To simplify our discussion, we assume the phase transition is first-order. Thus during the transition, bubbles of the leptonically CP-violating, broken phase begin to nucleate and expand within the symmetric phase. At a fixed space point around the bubble wall, $\lambda_{\alpha\beta}$ is time-dependent, $\lambda_{\alpha\beta}(t_{1})\neq\lambda_{\alpha\beta}(t_{2})$. Subsequently,  the lepton asymmetry is generated via the interference of the Weinberg operator at different times.

%%%%%%%%%%%%%%%%%%%%%%%%%%%%%%%%%%%%%%%%%%%%%%%%%%%%%%%%%%%%%%%%%%%%%%%%%%%%%%

\section{Closed-Time-Path Formalism}
Typically, observables in the thermal bath are derived from expectation values of operators that are not time-ordered. They can be calculated directly in the real time formalism, also known as the CTP formalism \cite{Schwinger:1960qe, Keldysh:1964ud}, which is derived from the first principles of Quantum Field Theory \cite{Calzetta:1986cq, Chou:1984es, Berges:2004yj}. This method has successfully been applied to leptogenesis based on the decay of right-handed neutrinos (see, e.g. \cite{Anisimov:2010dk, Beneke:2010wd, Beneke:2010dz}). 
In this letter, we  apply the CTP formalism to calculate the CPPT-induced lepton asymmetry. In comparison with   classical Boltzmann transport equations, in which the collision terms are calculated in zero temperature, an advantage of this formalism is the proper inclusion of quantum memory effects \cite{Anisimov:2010dk}.  As will be demonstrated, this effect plays a crucial role in our mechanism. 

For the Higgs ($H$) in the CTP formalism, one defines the following four Green functions:
\begin{eqnarray}
 \Delta^{T,\overline{T}}(x_1,x_2) &=& \langle T[H(x_1) H^*(x_2)] \rangle, \langle \overline{T}[H(x_1) H^*(x_2)] \rangle \,,\nonumber\\
 \Delta^{<,>}(x_1,x_2) &=& ~~ \langle H^*(x_2) H(x_1) \rangle,~~~~\langle H(x_1) H^*(x_2) \rangle \,,
 \label{eq:scalar_propagator}
\end{eqnarray}
where $T$ ($\overline{T}$) denotes time (anti-time) ordering. The Feynman, Dyson and Wightman propagators are represented by  $\Delta^{T}$, $\Delta^{\overline{T}}$ and $\Delta^{<,>}$, respectively.
Analogously, the definition of Green functions for  lepton ($\ell_\alpha$) with flavour index $\alpha$ is
\begin{eqnarray}
 S^{T,\overline{T}}_{\alpha\beta}(x_1,x_2) &=& \langle T[\ell_\alpha(x_1) \overline{\ell}_\beta(x_2)] \rangle, \langle \overline{T}[\ell_\alpha(x_1) \overline{\ell}_\beta(x_2)] \rangle \,,\nonumber\\
 S^{<,>}_{\alpha\beta}(x_1,x_2) &=& -\langle \overline{\ell}_\beta(x_2) \ell_\alpha(x_1) \rangle , ~~~\langle \ell_\alpha(x_1) \overline{\ell}_\beta(x_2) \rangle \,.
 \label{eq:fermion_propagator}
\end{eqnarray}
The additional minus sign in $ S^<$ comes from the anti-commutation property of fermions. In \equasref{eq:scalar_propagator}{eq:fermion_propagator}, electroweak gauge and fermion spinor indices have been suppressed. 

The Wightman propagators $S^{<,>}$ can be used to define the lepton asymmetry, e.g., the  number density difference between lepton and anti-lepton $n_{L}\equiv \sum_\alpha (n_{\ell\alpha}-n_{\overline{\ell}\alpha})$ 
\begin{eqnarray}
n_L(x) & = & - \frac{1}{2} \sum_\alpha \text{tr}\Big\{  \gamma^0 \big[S^<_{\alpha\alpha}(x,x) + S^>_{\alpha\alpha}(x,x) \big] \Big\} .
\label{eq:current}
\end{eqnarray}
Moreover, they satisfy the Kadanoff-Baym (KB) equation. We follow the convention of \cite{Prokopec:2003pj, Prokopec:2004ic, Garbrecht:2008cb} and express the KB equation as
\begin{eqnarray}
i \partial\!\!\!/ S^{<,>} - \Sigma^{H} \odot S^{<,>} - \Sigma^{<,>} \odot S^{H}  
= \frac{1}{2} \big[\Sigma^{>} \odot S^{<} - \Sigma^{<} \odot S^{>} \big] \,,
\label{eq:KB}
\end{eqnarray}
where $\odot$ denotes a convolution, $\Sigma$ is the self energy of the lepton, and $S^H$ and $\Sigma^H$ are Hermitian parts of propagator and self energy given by $S^H=S^T-\frac{1}{2}(S^>+S^<)$, $\Sigma^H=\Sigma^T-\frac{1}{2}(\Sigma^>+\Sigma^<)$ respectively. On the LHS of \equaref{eq:KB}, $\Sigma^{H}S^{<,>}$ represents the self-energy contribution and $\Sigma^{<,>} S^{H}$ describes the  broadening of the on-shell dispersion relation. On the RHS, $\frac{1}{2} (\Sigma^{>} S^{<} - \Sigma^{<} S^{>})$ is the collision term which includes the CP-violating source \cite{Prokopec:2003pj}.  As we  focus on the generation of an initial asymmetry, we consider only the collision term.

%%%%%%%%%%%%%%%%%%%%%%%%%%%%%%%%%%%%%%%%%%%%%%%%%%%%%%%%%%%%%%%%%%%%%%%%%%%%%%

\begin{figure}[h!]
\begin{center}
\includegraphics[scale=0.6]{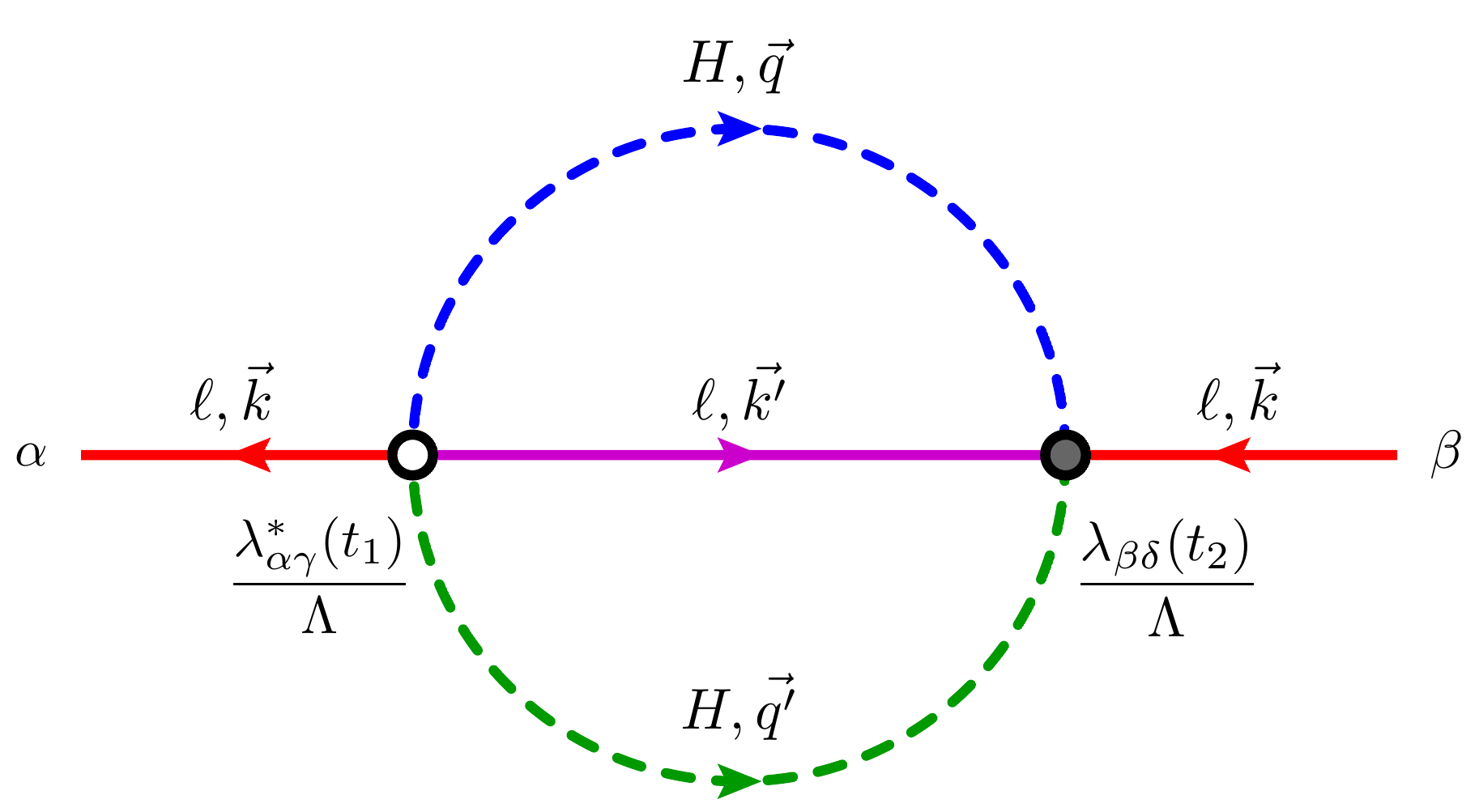}
\caption{The 2-loop lepton-number-violating contribution of the CP-violating and time-dependent Weinberg operator to the lepton self energy. }
\label{fig:Feynman}
\end{center}
\end{figure}

\section{Lepton Asymmetry} 
We follow the techniques developed for  thermal leptogenesis as presented in \cite{Anisimov:2010dk} and calculate the lepton asymmetry to leading order in a time-independent flavour basis. In order to derive  the lepton asymmetry,  the Green functions for the Higgs and leptons are Fourier transformed
\begin{eqnarray}
\Delta_{\vec{q}}(t_1,t_2) &=& \int d^3 r e^{i \vec{q} \cdot \vec{r}} \Delta(x_1,x_2)\,,\nonumber\\
S_{\vec{k}}(t_1,t_2) &=& \int d^3 r e^{i \vec{k} \cdot \vec{r}} S(x_1,x_2)\,,
\end{eqnarray}
where $r\equiv x_1-x_2$, $t_1\equiv x^0_1$ and $t_2\equiv x^0_2$. Subsequently, the lepton asymmetry at a fixed space point in the bubble wall may be written as $n_L(x) = \int \frac{d^3 k}{(2\pi)^3} L_{\vec{k}}$ with
\begin{eqnarray} 
L_{\vec{k}} \equiv f_{\ell\vec{k}} - f_{\overline{\ell}\vec{k}}  
&=& - \int_{t_i}^{t_f} dt_1 \partial_{t_1} \text{tr}[\gamma_0 S^<_{\vec{k}}(t_1,t_1) + \gamma_0 S^>_{\vec{k}}(t_1,t_1)] \nonumber\\
	 &=&  -\int_{t_{i}}^{t_{f}} dt_{1}\int_{t_{i}}^{t_{f}} dt_{2} \text{tr} \Big[\Sigma^{>}_{\vec{k}}(t_{1},t_{2})S_{\vec{k}}^{<}(t_{2},t_{1})  
	  -\Sigma^{<}_{\vec{k}}(t_{1},t_{2})S_{\vec{k}}^{>}(t_{2},t_{1})  \Big] \,,
 \label{eq:key}
\end{eqnarray}
where $t_{i}$ ($t_{f}$) is the initial (final) time and $\Sigma_{\vec{k}}(t_{1},t_{2})$ is the self-energy contribution. In this mechanism, the leading CP-violating contribution to $\Sigma_{\vec{k}}(t_{1},t_{2})$ is a 2-loop diagram as shown in \figref{fig:Feynman}. The memory effect is reflected in the `memory integral' over $t_1$ and $t_2$, which involves the time-dependent couplings shown in \figref{fig:Feynman}.  
Using \equaref{eq:Weinberg}, the lepton asymmetry may be re-expressed as
\begin{eqnarray}
L_{\vec{k}\alpha\beta} &=& \sum_{\gamma\delta}\frac{12}{\Lambda^{2}}\int_{t_{i}}^{t_{f}} dt_{1}\int_{t_{i}}^{t_{f}} dt_{2} \text{Im}\big\{ \lambda^{*}_{\alpha\gamma}(t_{1})\lambda_{\beta						\delta}(t_{2})\big\} 
\int_{q, q^{\prime}} M_{\alpha\beta\gamma\delta}(t_1,t_2,k,k^{\prime},q,q^{\prime}) \,,
\end{eqnarray}
where $
\int_{q, q^{\prime}}=\int \frac{d^{3}q}{(2\pi)^3} \frac{d^{3}q^{\prime}}{(2\pi)^3}$ 
and  $M_{\alpha\beta\gamma\delta}(t_1,t_2,k,k^{\prime},q,q^{\prime})$ is the finite temperature matrix element. 
Ignoring the differing flavours of lepton propagators and setting $t_i\to -\infty$, $t_f\to +\infty$, we obtain the total lepton asymmetry $L_{\vec{k}} \equiv \sum_{\alpha} L_{\vec{k}\alpha\alpha}$
\begin{eqnarray}
L_{\vec{k}} = \frac{12}{\Lambda^{2}} \int_{-\infty}^{+\infty} \!\!\!\!dt_{1}\int_{-\infty}^{+\infty} \!\!\!\!dt_{2}  \text{Im}\big\{ \text{tr} \left[ \lambda^{*}(t_{1})\lambda(t_{2})\right]\big\} \int_{q, q^{\prime}} M \,,
\end{eqnarray}
where the  finite temperature matrix element, decomposed in terms of the lepton and Higgs propagators, is given by
\begin{eqnarray}\label{eq:ME}
M = \text{Im}\big\{ \Delta^{<}_{\vec{q}}(t_1,t_2)  \Delta^{<}_{\vec{q^{\prime}}}(t_1,t_2) \text{tr} \big[ S^{<}_{\vec{k}}(t_1,t_2)S^{<}_{\vec{k^{\prime}}}(t_1,t_2) P_{L}  \big]   \big\}. 
\end{eqnarray}

A homogeneous system is a principle assumption in the derivation of \equaref{eq:key} \cite{Anisimov:2010dk}. However, this is clearly not the case for CPPT as the bubble expansion provides a special direction perpendicular to the bubble wall which results in the transport of the lepton asymmetry along this particular direction.  We anticipate the directional dependence of the asymmetry will be small and therefore ignore its impact at this stage.  As  the temperature at which CPPT occurs is  significantly higher than the electroweak scale, both leptons and the Higgs are almost in thermal equilibrium. Using this assumption, we prove in the Appendix  the lepton asymmetry from spatial-dependent EEV profile vanishes. We shall apply this approximation throughout this work. 

The time-dependent flavon EEV, $\langle\phi\rangle$, plays an important role in CPPT.  Without loss of generality, one may assume $\langle\phi\rangle = v_{\phi}f(t)$ where $f(t)$ varies continuously from $0$ to $1$. The coupling coefficient, $\lambda$, takes the form
\begin{eqnarray}
\lambda(t) =\lambda^0+ \lambda^1 f(t)\,.
\label{eq:vev_profile}
\end{eqnarray}  
The lepton asymmetry should not be sensitive to the precise functional form of  $f(x)$. The simplest example of $f(t)$ is a step function, $f(t)= \vartheta(v_w t - z)$, where $v_{w}$ is the velocity of the bubble wall and $z$ a certain point along the direction of bubble expansion. Another example is a tanh function $f(t) = \frac{1}{2}\Big[ 1+\tanh \left( \frac{v_{w}t-z}{L_{w}}  \right)  \Big]$
where $L_{w}$ the width of the wall. The latter case is analogous to the Higgs EEV profile studied in the electroweak strong first-order phase transition, which has been numerically checked \cite{Moore:1995ua, Moore:1995si, Cline:2011mm, Konstandin:2014zta, Kozaczuk:2015owa}. In the limit $L_w\to 0$, the second example reduces to the first. Both cases yield the same result, shown in  \equaref{eq:integration}, as expected. In addition to a first-order phase transition, it is possible a a sudden
change in the VEV of the scalar field, $\phi$,
may be generated by dynamics other  than a thermal first-order phase
transition such as a quench  in the context of
cold electroweak baryogenesis \cite{GarciaBellido:1999sv,Krauss:1999ng}
.
Then $\text{Im}\big\{ \text{tr}\left[\lambda^{*}(t_{1})\lambda(t_{2})\right]\big\}$ may be rewritten as
\begin{eqnarray}
 \, \text{Im}\big\{ \text{tr}\left[\lambda^{*}(t_{1})\lambda(t_{2})\right]\big\} = \text{Im}\big\{  \text{tr}\left[\lambda^{0}\lambda^*\right] \big\} \left[f\left(t_1\right) - f\left(t_2\right)  \right] \,. 
\label{eq:integration}
\end{eqnarray}
After an exchange of integration variables from $t_1$ ($t_2$) to $\tilde{t}=(t_1+t_2)/2$ ($y=t_1-t_2$) and using
\begin{eqnarray}\int_{-\infty}^{+\infty} d\tilde{t} [f(\tilde{t}+y/2)-f(\tilde{t}-y/2)] = y,
\end{eqnarray}
we integrate $\tilde{t}$ to find
\begin{eqnarray}
\!\!\int_{-\infty}^{+\infty} d\tilde{t} \, \text{Im}\big\{ \text{tr}\left[\lambda^{*}(t_{1})\lambda(t_{2})\right]\big\} = \text{Im}\big\{  \text{tr}\left[\lambda^{0}\lambda^*\right] \big\} y \,. 
\label{eq:integration_v2}
\end{eqnarray}
Hence the lepton asymmetry becomes
\begin{eqnarray}\label{eq:asymmetry}
L_{\vec{k}} 
=  - \frac{12}{v^{4}_{H}} \text{Im}\{ \text{tr}[ m^{0}_{\nu}m_{\nu}^* ]\} \int_{-\infty}^{+\infty}dy y \int_{q,q^{\prime}} M \,,
\end{eqnarray}
where we have reparametrised the effective neutrino mass matrices as $m_\nu^0 \equiv \lambda^0 v_H^2/\Lambda$ and $m_\nu  \equiv (\lambda^0 + \lambda^1) v_H^2/\Lambda$. As $v_w$ and $L_w$ have dropped out of \equaref{eq:asymmetry}, the bubble wall properties do not affect the final result of the lepton asymmetry.  This is based on the assumptions of single-scalar phase transition and fast bubble expansion. \equaref{eq:integration} is not valid for an arbitrary $\lambda(t)$ shape. Instead, if the interference time scale is much smaller than the wall expansion scale, we can perturb $y$ and obtain 
\begin{eqnarray}
\text{Im}\big\{ \text{tr}\left[\lambda^{*}(t_{1})\lambda(t_{2})\right]\big\} \approx \text{Im}\big\{  \text{tr}\left[\lambda^*(\tilde{t})\partial_{\tilde{t}}\lambda(\tilde{t})\right] \big\} y \,. 
\label{eq:integration_v3}
\end{eqnarray}
On the other hand, the temperature evolution should be considered if the bubble expands not fast enough compared with the Hubble expansion. 
These cases will be discussed in detail in our future work \cite{future}.  
In the remainder of this work, we shall still continue to work in these assumptions. 

To evaluate \equaref{eq:asymmetry}, firstly we  calculate $M$ using the lepton and Higgs propagators.  As the scale of the CPPT is significantly higher than that of the electroweak scale, we will assume that leptons and Higgs are in thermal equilibrium. In the massless limit, the lepton and Higgs propagators are written as
\begin{eqnarray}\label{eq:props}
  \Delta^{<}_{\vec{q}}(t_1,t_2) &=& \frac{\text{c}_q}{2q\, \text{sh}_q}e^{-\gamma_{H}\left| y \right|} \,,\nonumber\\    
         S^{<}_{\vec{k}}(t_1,t_2) &=&  \frac{\gamma^{0}\text{c}_k +i\vec{\gamma}\cdot\hat{k}\,\text{s}_k}{2\,\text{ch}_k}e^{-\gamma_{\ell} \left| y \right|} \,,
\end{eqnarray}
where $k = |\vec{k}|$, $q = |\vec{q}|$, $\text{c}_k=\cos(ky^-)$, $\text{s}_k=\sin(ky^-)$, $\text{ch}_k=\cosh(k\beta/2)$, $\text{sh}_k=\sinh(k\beta/2)$, $y^{-}=y-i\beta/2$, $\hat{k}=\vec{k}/k$ and $\gamma_{H}$, $\gamma_{\ell}$ are the thermal damping rates of the Higgs and the leptonic doublets  respectively  \cite{Anisimov:2010dk}. 
Substituting \equaref{eq:props} into \equaref{eq:ME}, $M$ becomes
\begin{eqnarray}
M=   \frac{\text{Im} \big\{\text{c}_q \text{c}_{q^{\prime}} [\text{c}_k \text{c}_{k^{\prime}} + \hat{k}\cdot\hat{k^{\prime}} \text{s}_k \text{s}_{k^{\prime}} ] \big\} }{8qq^{\prime}\text{sh}_q \text{sh}_{q^{\prime}} \text{ch}_k \text{ch}_{k^{\prime}} } e^{-2\gamma \left| y\right|} \,,
\end{eqnarray}
where $\gamma=\gamma_{\ell}+\gamma_{H}$. In order to simplify \equaref{eq:asymmetry}, we will first perform the integration of the $y$ variable and then  the momentum integration. The $y$ integration may be performed by exploiting that  $M$ is an odd function of $y$
\begin{eqnarray}\label{eq:ME2}
\int_{-\infty}^{+\infty} dy y M =2\int_{0}^{\infty}dy y M 
=-\!\!\!\!\!\sum_{\eta_{2},\eta_{3},\eta_{4}=\pm1}\frac{K_{\eta_2\eta_3\eta_4}\gamma\sin\left(\beta K/2\right)[1-\eta_{2}\hat{k}\cdot\hat{k^{\prime}}]}{16qq^{\prime}\left(K_{\eta_2\eta_3\eta_4}^2+4\gamma^{2}\right)^{2}\text{sh}_q \text{sh}_{q^{\prime}} \text{ch}_k \text{ch}_{k^{\prime}} } \,,
\end{eqnarray}
where $K_{\eta_2\eta_3\eta_4}=k+\eta_{2}k^{\prime}+\eta_{3}q+\eta_{4} q^{\prime}$.

The evaluation of the momentum integration will closely follow that of \cite{Anisimov:2010dk}. We abstain from re-deriving the details of this calculation and instead refer the reader to the reference. However we will present the simplified form of the momentum integration
\begin{eqnarray}\label{eq:momentum}
\int_{q, q^{\prime}} = \frac{1}{\left(2\pi\right)^{4}} \int_{0}^{\infty}\!\!\!\!dp\int_{0}^{\infty}\!\!\!\!k^{\prime}dk^{\prime} \int_{\left| k-p \right|}^{\left| k+p \right|}\!\!\!\!qdq \int_{\left| k^{\prime}-p \right|}^{\left| k^{\prime}+p \right|}\!\!\!\!q^{\prime}dq^{\prime} \,,
\end{eqnarray}
where $ p = k-q = k^{\prime}-q^{\prime}$, $q^{2} = k^{2}+p^{2}-2pk\cos\theta$ and ${q^{\prime}}^{2} = {k^{\prime}}^{2}+p^{2}-2pk^{\prime}\cos\theta $
have been applied.  Using \equaref{eq:ME2} together with \equaref{eq:momentum}, the final result is written as
\begin{eqnarray}
L_{\vec{k}} = \frac{3\,\text{Im}\Big\{ \text{tr}\left[ m^{0}_{\nu}m^{*}_{\nu} \right]  \Big\}T^2}{\left( 2\pi \right)^{4}v^{4}_{H}}  F\left( x_{1},x_{\gamma} \right)\,.
\end{eqnarray}
 $F\left( x_{1},x_{\gamma} \right)$ is a loop factor given by
\begin{eqnarray}
F\left( x_{1},x_{\gamma} \right) &=& \frac{1}{x_{1}} \int_{0}^{+\infty}\!\!\!\!dx  \int_{0}^{+\infty}\!\!\!\!\!x_{2}dx_{2} \int_{\left|x_{1}-x \right|}^{\left|x_{1}+x \right|}\!\!\!\!dx_{3}  \int_{\left|x_{2}-x \right|}^{\left|x_{2}+x \right|}\!\!\!\!dx_{4} 
\sum_{\eta_{2},\eta_{3},\eta_{4}=\pm1} \bigg[1-\frac{\left(x^2_{1}+x^{2}-x^{2}_3\right)\left(x^2_{2}+x^{2}-x^{2}_4\right)}{4\eta_{2}x_{1}x_{2}x^2} \bigg]\nonumber\\
&&\times \frac{X_{\eta_{2}\eta_{3}\eta_{4}}x_{\gamma}\sinh X_{\eta_{2}\eta_{3}\eta_{4}}}{\left(X^2_{\eta_{2}\eta_{3}\eta_{4}}+x^2_{\gamma}\right)^{2}\cosh x_{1}\cosh x_{2}\sinh x_{3}\sinh x_{4}} \,,
\end{eqnarray}
where $x_{1}=k\beta/2$,  $x_{2}=k^{\prime}\beta/2$,  $x_{3}=q\beta/2$,  $x_{4}=q^{\prime}\beta/2$,  $x=p\beta/2$, $x_\gamma=(\gamma_\ell+\gamma_H)\beta$ and  $X_{\eta_{2}\eta_{3}\eta_{4}}=x_{1}+\eta_{2}x_{2}+\eta_{3}x_{3}+\eta_{4}x_{4}$. 
The loop factor is dependent upon the lepton energy and the thermal width normalised by the temperature, i.e., $x_1$ and $x_\gamma$. 
Unlike conventional scenarios of leptogenesis, this mechanism has the feature that once the  lepton asymmetry is produced, it will not be washed out. 
This is because the lepton anti-lepton rate proceeds via the Weinberg operator which may be approximated as
\begin{eqnarray}
\Gamma_{W}  \simeq \frac{3}{4\pi^3}\frac{\lambda^2}{\Lambda^2}T^3 \simeq \frac{3}{4\pi^3}\frac{m_\nu^2}{v_H^4} T^3\,.
\end{eqnarray}
We shall find the rate of the Weinberg operator mediated washout will be significantly less than the Hubble expansion rate at the temperature of the phase transition. Thus, the washout processes via the Weinberg operator are not active. 
The lepton asymmetry is partially converted into the baryon asymmetry via electroweak sphaleron processes which are active above the electroweak scale. However, $n_{B-L} \equiv - n_L(T=T_\text{CPPT})$ is conserved, where $T_\text{CPPT}$ is the CPPT temperature. The final baryon symmetry is roughly given by $n_B \approx \frac{1}{3} n_{B-L}$. The baryon-to-photon ratio $\eta_B$ is defined as
\begin{eqnarray} 
\eta_B\equiv \frac{n_B}{n_\gamma} \approx - \frac{\text{Im}\{\text{tr}[m_\nu^{0} m_\nu^*] \} T^2}{8 \pi^4 \zeta(3) v_H^4 } F(x_\gamma) \,,
\end{eqnarray}
where $F(x_\gamma)=\int_0^{+\infty} x_1 dx_1 F(x_1, x_\gamma)$, $n_\gamma= 2 \zeta(3) T^3/\pi^2$ and $\zeta (3) = 1.202$. 
In order to produce a positive baryon asymmetry, $\text{Im}\{\text{tr}[m_\nu^{0} m_\nu^*] \}$ should take a minus sign. Importantly, the lepton asymmetry is independent of the choice of flavour basis. 

\section{Discussion} 
The lepton asymmetry, as shown in \equaref{eq:key}, is  dependent upon three components:  the loop factor $F(x_1, x_\gamma)$ derived from the correction to the lepton propagator; the effective neutrino mass matrices $m_\nu^0$, $m_\nu$  and  the temperature, $T$, at which CPPT occurs. We shall address each of these contributions in turn.

\begin{figure}[h!]
\begin{center}
\includegraphics[scale=0.6]{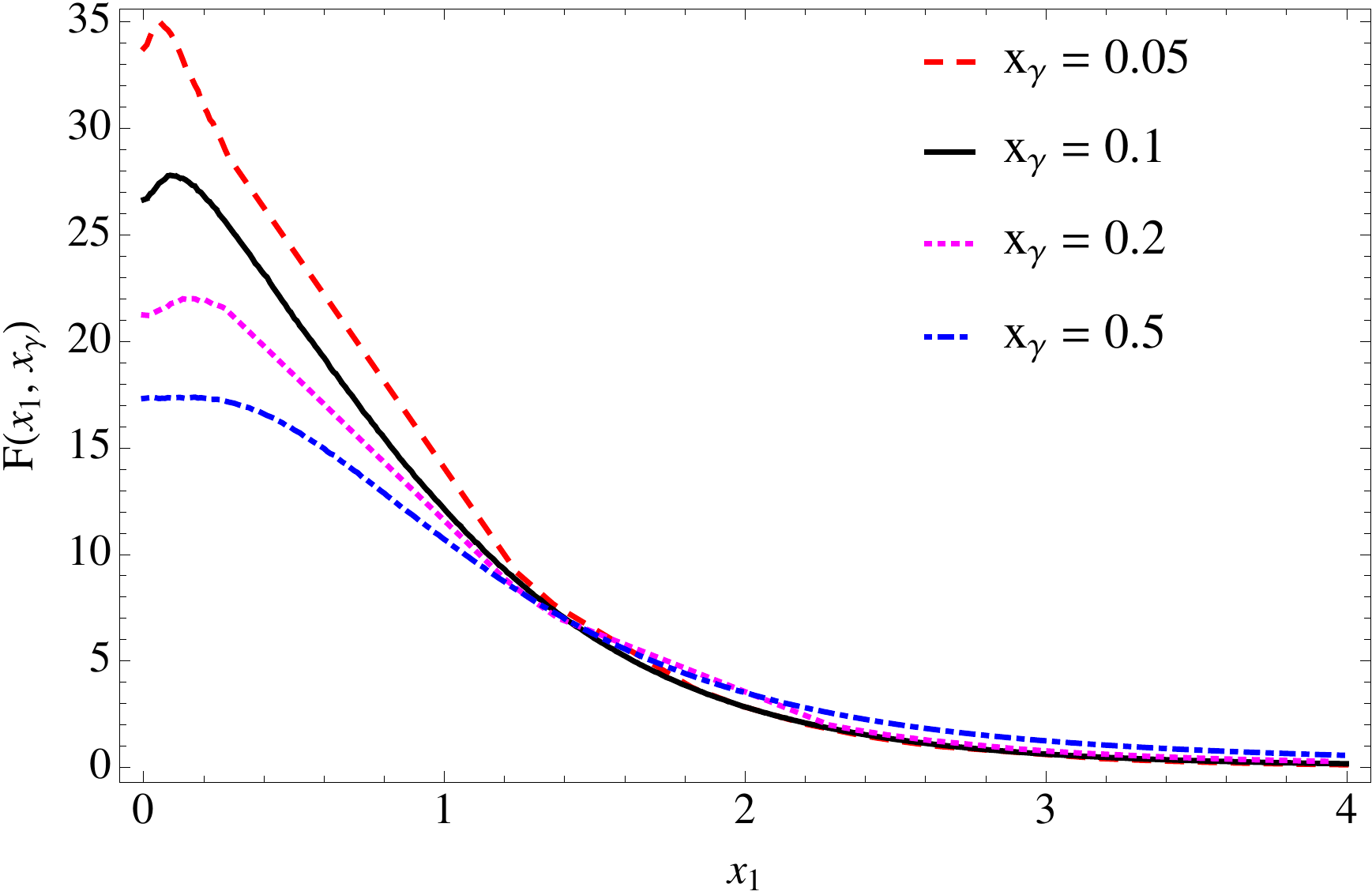}
\caption{The loop factor $F(x_1,x_\gamma)$ as a function of the lepton energy $k$ and the thermal width $\gamma$, where $x_1 = k\beta/2$ and $x_\gamma = \gamma \beta$. }
\label{fig:loopfactor}
\end{center}
\end{figure}

Thermal widths play important role in the mechanism. It corresponds to the decoherence of the Weinberg operators at large time difference. In \figref{fig:loopfactor}, we allow $x_\gamma$ to vary and display the numerical results of the loop factor $F(x_1,x_\gamma)$ as a function of $x_1$. As expected, a smaller thermal width results in a larger lepton asymmetry since the interference with larger time difference $|y|$ is less suppressed. The main contribution to the Standard Model value, $x_\gamma \approx 0.1$, mainly comes from electroweak gauge couplings \cite{bellac}. High scale new physics may enlarge $x_\gamma$.  For $x_1 \sim 1$, we observe the loop factor provides a factor $\mathcal{O}(10)$ enhancement to the lepton asymmetry. 

We have introduced the effective neutrino mass matrices $m_\nu^0$ and $m_\nu$.  The structure of $m_\nu^0$  is dependent on the coupling of the flavons to the Weinberg operator. The form of this coupling is determined by the details of particular flavour models and will be studied elsewhere. After CPPT, the coefficients of the Weinberg operator are fixed and  $m_\nu$ is identified to the measurable low-energy neutrino mass matrix (ignoring RG running effects). 
This mass matrix is diagonalised by the PMNS matrix, i.e.,$U_\text{PMNS}^T m_\nu U_\text{PMNS}^*=\text{diag}\{m_1,m_2,m_3\}$, which allows for a  connection between lepton asymmetry and low-energy leptonic observables. As discussed in Section (4), the washout is proportional to the light neutrino mass matrix. For the temperatures, $T\sim 10^{11}$ GeV, the washout rate is much less than the Hubble expansion rate and therefore these processes are not effective during the phase transition. 
However, the temperature of the phase transition may be higher or lower than $10^{11}$ GeV, depending on the value of $m^0_{\nu}$. The washout rate provides an upper bound for the phase transition as these rates freeze out around $T_{B}\sim 10^{13}$ GeV. Therefore, if the CP-violating phase transition occurred above $T_B$ the generated lepton asymmetry would be completely washed out. 

Finally, we discuss the associated energy scale of CPPT. 
In order to estimate the temperature at which CPPT occurs, we will assume $\text{Im}\{\text{tr}[m_\nu^{0} m_\nu^*]\}$ is  the same order as $m_\nu^2 \sim (0.1~\text{eV})^2$. Numerically, we have checked that $F(x_\gamma)$ provides an $\mathcal{O}(10^2)$ factor enhancement for $x_\gamma \sim 0.1$. 
Thus, the temperature for successful CPPT is 
\begin{eqnarray}
T_\text{CPPT} \sim \sqrt{\eta_{B}}\, \frac{v_H^2}{m_\nu} \,.
\end{eqnarray} 
Using the observed ratio of baryon to photon,  $\eta_B=(6.19\pm0.15) \times10^{-10}$ \cite{PDG},  we conclude $T_\text{CPPT} \sim 10^{11}$ GeV. This is a rough estimate and a more detailed calculation, involving the evolution of the initial asymmetry and inclusion of effects such as differing thermal width of charged leptons, may lower this scale. 
This mechanism is contingent upon the UV-completion scale $\Lambda$ being higher than the temperature of phase transition $T$. If $\Lambda\lesssim T$, new lepton-number-violating particles, e.g., right-handed neutrinos in type-I seesaw, may be involved in the thermal bath during the phase transition, and the phase transition may influence the leptogenesis via the decays of these particles \cite{Pilaftsis:2008qt}. 

There are several ways to further refine the calculation of  the  lepton asymmetry generated from CPPT. Firstly,
we neglected the thermal masses of the Higgs and leptons. Proper inclusion of these masses would have the effect of mild suppression of the loop function. In addition, the thermal width we applied ($x_{\gamma}\sim 0.1$) is an  effective description which estimates their effects. A more complete treatment would involve the explicit calculation of the imaginary part of the self-energy correction to the Higgs and lepton propagators. 
In the above, we have only calculated the time component lepton asymmetry generated by the time-dependent coupling of the Weinberg operator. The coupling may also be space-dependent during the phase transition.  We comment in the rest frame of the plasma, that the spatial component lepton asymmetry is negligibly small. This is due to our assumption of thermal equilibrium of the propagators, c.f. Eq. (15). Therefore, the momentum distributions of the Higgs and leptons are spatially isotropic. Consequently, there is no preferred direction for the Higgs and lepton propagators, and therefore, combining these propagators with the space-dependent coupling, which specifies the $z$ direction, cannot generate any lepton asymmetry \cite{Lee:2004we}. It is in principle possible that a deviation from the equilibrium may result in additional lepton asymmetry nonetheless this contribution can be safely ignored for temperatures  far above the electroweak scale as the deviation is very small and the generated lepton asymmetry would be negligible. 

\section{Conclusion} 

We have proposed a novel mechanism based on CPPT to generate  the matter-antimatter asymmetry. It differs from conventional high-scale leptogenesis scenarios as we assume CP is broken below the scale of neutrino mass generation and apply an effective theory approach which permits model independence. Moreover, this mechanism allows for a connection between leptonic flavour structure and the baryon asymmetry. 

The essential requirements of this approach are a CP-violating phase transition and the Weinberg operator. We assume the complex coefficients of this operator are dynamically realised. During the phase transition, the lepton asymmetry is generated from the interference of the Weinberg operator at different times. In order to generate the observed baryon asymmetry, the temperature scale of  CPPT is  approximately $T_\text{CPPT} \sim 10^{11}$ GeV.

\section*{Acknowledgements} 

It is a pleasure to thank Marco Drewes for helpful discussions on the CTP formalism. We are grateful to Carlos Tamarit and Alexis Plascencia for fruitful discussions on electroweak baryogenesis and leptogenesis. We would also like to thank Nicholas Jennings and Shun Zhou for useful feedback and  Alan Reynolds for his insightful question on the functional form of the bubble wall. This work has been supported by the European Research Council under ERC Grant “NuMass” (FP7-IDEAS-ERC ERC-CG 617143), H2020 funded ELUSIVES ITN (H2020-MSCA-ITN-2015, GA-2015-674896-ELUSIVES), InvisiblePlus (H2020-MSCA-RISE-2015, GA-2015-690575-InvisiblesPlus) and the Science and Technology Facilities Council (STFC).

\appendix
\section{}

We explicitly provide a proof below which demonstrates that the spatial contribution to the 
lepton asymmetry is negligible.

We start from the following formula of the lepton number asymmetry $N_L \equiv N_\ell - N_{\overline{\ell}}$ in the CTP approach:  
\begin{eqnarray} 
N_L 
&=& - \int d^4x_1 d^4x_2 \text{tr}\Big[\Sigma^{>}(x_1,x_2)  S^{<}(x_2,x_1) - \Sigma^{<}(x_1,x_2)  S^{>}(x_2,x_1) \Big] \,,
\end{eqnarray}
which is derived from the Kadanoff-Baym equation.  Including the diagram in Fig. 1 in our paper, we obtain 
\begin{eqnarray}
N_L &=& - \frac{12}{\Lambda^2} \int d^4 \tilde{x} d^4r\text{Im}\big\{\text{tr} [\lambda^*(\tilde{x}+\frac{r}{2}) \lambda(\tilde{x}-\frac{r}{2})] \big\} \mathcal{M}\,,
\end{eqnarray}
where $\tilde{x}=(x_1+x_2)/2$, $r=x_1-x_2$ and 
\begin{eqnarray}
\mathcal{M} &=& \text{Im} \Big\{ \int\frac{d^4k}{(2\pi)^4} \frac{d^4k'}{(2\pi)^4} \frac{d^4q}{(2\pi)^4} \frac{d^4q'}{(2\pi)^4}  e^{i(k+k'+q+q')\cdot r} \Big[ \Delta^{<}_{q} \Delta^{<}_{q'} \text{tr} [ S^{<}_{k} S^{<}_{k'} ] - \Delta^{>}_{q} \Delta^{>}_{q'} \text{tr} [ S^{>}_{k} S^{>}_{k'} ] \Big] \Big\} \,.
\end{eqnarray} 
The Wightman propagators $\Delta^{<,>}_{q}$ and $S^{<,>}_{k}$ at tree level are given by
\begin{eqnarray}
\Delta^<_{q}(x) &=& 2\pi \delta(q^2) \Big\{ \vartheta(q_0) f_{H, \vec{q}} (x) + \vartheta(-q_0) [1+ f_{H^*, -\vec{q}}(x) ] \Big\} \,, \nonumber\\
\Delta^>_{q}(x) &=& 2\pi \delta(q^2) \Big\{ \vartheta(q_0) [1+f_{H,\vec{q}} (x)] + \vartheta(-q_0) f_{H^*,-\vec{q}}(x) \Big\}, \,\nonumber\\
S^<_{k}(x) &=& -2\pi \delta(k^2) P_L k\!\!\!/ P_R \Big\{ + \vartheta(k_0) f_{\ell, \vec{k}} (x) - \vartheta(-k_0)[1- f_{\overline{\ell}, -\vec{k}}(x) ] \Big\} \,, \nonumber\\
S^>_{k}(x) &=& -2\pi \delta(k^2) P_L k\!\!\!/ P_R \Big\{ - \vartheta(k_0) [1-f_{\ell,\vec{k}} (x)] + \vartheta(-k_0) f_{\overline{\ell}, -\vec{k}}(x) \Big\},
\label{eq:propagator}
\end{eqnarray}
with $f_{H, \vec{q}} (x)$, $f_{H^*, \vec{q}} (x)$,  $f_{\ell, \vec{k}} (x)$ and $f_{\overline{\ell}, \vec{k}} (x)$ the distribution densities of $H$, $H^*$, $\ell$ and $\overline{\ell}$ respectively. 

By assuming the one scalar case in \equaref{eq:vev_profile} and applying \equaref{eq:integration} of our paper, we first integrate over $\tilde{t}\equiv \tilde{x}^0$  and obtain Eq. (13), i.e., 
\begin{eqnarray}
\!\!\int_{-\infty}^{+\infty} d\tilde{t} \, \text{Im}\big\{ \text{tr}\left[\lambda^{*}(\tilde{t}+\frac{y}{2})\lambda(\tilde{t}-\frac{y}{2})\right]\big\} = \text{Im}\big\{  \text{tr}\left[\lambda^{0}\lambda^*\right] \big\} y \,. 
\label{eq:integration_v4}
\end{eqnarray}
Including the velocity of the bubble wall can be done  by replacing $\tilde{t}\pm y/2$ with $(\tilde{x}^0 - \tilde{x}^3/v_w) \pm (r^0 - r^3/v_w)/2$, where $v_w$ is the velocity of the bubble wall. Following the same procedure, we integrate over $\tilde{x}$ first, and obtain 
\begin{eqnarray}
\int d^4\tilde{x} \text{Im}\big\{\text{tr} [\lambda^*(\tilde{x}+\frac{r}{2}) \lambda(\tilde{x}-\frac{r}{2})] \big\} &=& \text{Im}\{ \text{tr} [ \lambda^{0} \lambda^* ] \} V \, \Big(r^0-\frac{r^3}{v_w}\Big) \,,
\label{eq:int_x}
\end{eqnarray}
where $V = \int d^3 \vec{\tilde{x}}$ is a chosen sufficiently large volume in the space. This leads us to the  expression for the lepton number per unit volume $N_L/V = n_L + n_L'$, where
\begin{eqnarray} 
n_L \equiv \int \frac{d^3 k}{(2\pi)^3} L_{\vec{k}} &=& - \frac{12}{\Lambda^2} \text{Im}\{ \text{tr} [ \lambda^{0} \lambda^* ] \} \int d^4r\, r^0 \,\mathcal{M} \,, \nonumber\\
n_L' \equiv \int \frac{d^3 k}{(2\pi)^3} L_{\vec{k}}' &=& + \frac{12}{\Lambda^2} \text{Im}\{ \text{tr} [ \lambda^{0} \lambda^* ] \} \int d^4r\, \frac{r^3}{v_w} \mathcal{M}\,,
\label{eq:LeptonAsymm}
\end{eqnarray}
represent the time-dependent and space-dependent lepton asymmetry, corresponding to integrations along $r^0\equiv y$ and $r^3/v_w$, respectively. 

As the temperature is much higher than the EW scale, we assume Wightman propagators of the Higgs and leptons are in thermal equilibrium in the rest frame of the plasma, 
\begin{eqnarray}
&&f_{H, \vec{q}}(x) = f_{H^*, \vec{q}}(x) = f_{B,|q^0|} \equiv \frac{1}{e^{\beta |q^0|}-1} \,,\nonumber\\
&&f_{\ell, \vec{k}}(x) \;= f_{\overline{\ell}, \vec{k}}(x)\;\;\; = f_{F,|k^0|} \equiv \frac{1}{e^{\beta |k^0|}+1} \,.
\end{eqnarray}

We apply the following parity transformation for $\mathcal{M}$: 
\begin{eqnarray}
r \to r^P = (r^0,-\vec{r})\,, \quad k_n \to k_n^P=(k^0_n, -\vec{k}_n)\,,
\label{eq:parity}
\end{eqnarray} 
where $k_n$ represents each of $k,k'q,q'$. 
Note that $\Delta_q^{<,>}$ is invariant under the spatial parity transformation, $\Delta_q^{<,>}=\Delta_{q^P}^{<,>}$. Although $S^{<}_{k}$ is not invariant under $k \to k^P=(k^0,-\vec{k})$, but $ \text{tr} [ S^{<,>}_{k^P} S^{<,>}_{k^{\prime P}} ] =  \text{tr} [ S^{<,>}_{k} S^{<,>}_{k'} ]$. From these properties, we directly prove that $\mathcal{M}$ is invariant under the parity transformation in Eq. \eqref{eq:parity}. In other words, $\mathcal{M}$ is an even function of $\vec{r}$. 

~~ We highlight that $L_{\vec{k}}$  in Eq.~\eqref{eq:LeptonAsymm} is just as the same as that in our paper in \equaref{eq:integration_v2}. This is easily checked by performing the spatial integration $d^3\vec{r}$ in Eq.~\eqref{eq:LeptonAsymm}. The space-dependent integration $n_L'$ is zero. This is because $\mathcal{M}$ is an even function of $r^3$, and thus the space-dependent integration, $\int d^4r\, r^3 \mathcal{M}$, vanishes. 

The result in Eq.~\eqref{eq:LeptonAsymm} is based on the single-flavon case the only interference term between $\lambda^0$ and $\lambda^1f(x)$ can be integrated along $\tilde{x}$ directly. In the multi-scalar case, $\lambda = \lambda^0+\sum_i \lambda^i f_i(x)$, the interference between $\lambda^i f_i(x_1)$ and $\lambda^j f_j(x_2)$ cannot be integrated along $\tilde{x}$ directly as in Eq.~\eqref{eq:int_x}. Instead, we can do the ``thick wall'' approximation, where the interference length is much smaller than the wall thickness. In the case, we perturb $r$ and derive
\begin{eqnarray}
\int d^4r \text{Im}\{ \text{tr} [ \lambda^{*}(x+r/2) \lambda(x-r/2) ] \} \mathcal{M} &\approx& \text{Im}\{ \text{tr} [ \lambda^{*}(x) \partial_\mu \lambda(x) ] \} \int d^4r r^\mu  \mathcal{M} \,.
\label{eq:source_Lorentz}
\end{eqnarray}
It is useful to define the CP sources per unit volume per unit time $\mathcal{S}^{\CPV}_\ell(x)$ and $\mathcal{S}^{\prime\CPV}_\ell(x)$ as 
\begin{eqnarray}
\mathcal{S}^{\CPV}_\ell (x) &=& - \frac{12}{\Lambda^2} \text{Im}\{ \text{tr} [ \lambda^{*}(x) \partial_0 \lambda(x) ] \} \int d^4r r^0  \mathcal{M} \,,\nonumber\\
\mathcal{S}^{\prime\CPV}_\ell (x) &=& - \frac{12}{\Lambda^2} \text{Im}\{ \text{tr} [ \lambda^{*}(x) \partial_3 \lambda(x) ] \} \int d^4r r^3  \mathcal{M} \,.
\label{eq:source_v1}
\end{eqnarray}
Again, the space-dependent integration $\mathcal{S}^{\prime\CPV}_\ell(x)$ vanishes, and only the time-dependent integration remains. In summary, the spatial contributions
to the asymmetry may be safely neglected, both in the thin-wall and thick wall scenarios,  as we assume the Higgs and leptons are in thermal equilibrium which is a reasonable assumption at $T\sim10^{11}$ GeV.\\


\begin{thebibliography}{99}

\bibitem{Fukugita:1986hr} 
  M.~Fukugita and T.~Yanagida,
  %``Baryogenesis Without Grand Unification,''
  Phys.\ Lett.\ B {\bf 174} (1986) 45.
%  doi:10.1016/0370-2693(86)91126-3

\bibitem{Khlebnikov:1988sr} 
  S.~Y.~Khlebnikov and M.~E.~Shaposhnikov,
  %``The Statistical Theory of Anomalous Fermion Number Nonconservation,''
  Nucl.\ Phys.\ B {\bf 308} (1988) 885.
%  doi:10.1016/0550-3213(88)90133-2

\bibitem{Sakharov:1967dj} 
  A.~D.~Sakharov,
  %``Violation of CP Invariance, c Asymmetry, and Baryon Asymmetry of the Universe,''
  Pisma Zh.\ Eksp.\ Teor.\ Fiz.\  {\bf 5} (1967) 32
  [JETP Lett.\  {\bf 5}, 24 (1967)]
  [Sov.\ Phys.\ Usp.\  {\bf 34} (1991) 392]
  [Usp.\ Fiz.\ Nauk {\bf 161} (1991) 61].
%  doi:10.1070/PU1991v034n05ABEH002497

\bibitem{DUNE}
%\bibitem{Acciarri:2015uup} 
  R.~Acciarri {\it et al.} [DUNE Collaboration],
  %``Long-Baseline Neutrino Facility (LBNF) and Deep Underground Neutrino Experiment (DUNE) : Volume 2: The Physics Program for DUNE at LBNF,''
  arXiv:1512.06148 [physics.ins-det].

\bibitem{T2HK}
%\bibitem{Abe:2015zbg} 
  K.~Abe {\it et al.} [Hyper-Kamiokande Proto-Collaboration],
  %``Physics potential of a long-baseline neutrino oscillation experiment using a J-PARC neutrino beam and Hyper-Kamiokande,''
  PTEP {\bf 2015} (2015) 053C02
%  doi:10.1093/ptep/ptv061
  [arXiv:1502.05199 [hep-ex]].

\bibitem{Froggatt:1978nt} 
  C.~D.~Froggatt and H.~B.~Nielsen,
  %``Hierarchy of Quark Masses, Cabibbo Angles and CP Violation,''
  Nucl.\ Phys.\ B {\bf 147} (1979) 277.
%  doi:10.1016/0550-3213(79)90316-X

\bibitem{Alonso:2013nca} 
  R.~Alonso, M.~B.~Gavela, G.~Isidori and L.~Maiani,
  %``Neutrino Mixing and Masses from a Minimum Principle,''
  JHEP {\bf 1311} (2013) 187
  %doi:10.1007/JHEP11(2013)187
  [arXiv:1306.5927 [hep-ph]].

\bibitem{King:2014nza} 
  for a review, see e.g., S.~F.~King, A.~Merle, S.~Morisi, Y.~Shimizu and M.~Tanimoto,
  %``Neutrino Mass and Mixing: from Theory to Experiment,''
  New J.\ Phys.\  {\bf 16} (2014) 045018
%  doi:10.1088/1367-2630/16/4/045018
  [arXiv:1402.4271 [hep-ph]].

\bibitem{Branco:2011zb} 
  G.~C.~Branco, R.~G.~Felipe and F.~R.~Joaquim,
  %``Leptonic CP Violation,''
  Rev.\ Mod.\ Phys.\  {\bf 84} (2012) 515
%  doi:10.1103/RevModPhys.84.515
  [arXiv:1111.5332 [hep-ph]].

\bibitem{deMedeirosVarzielas:2011zw} 
  I.~de Medeiros Varzielas and D.~Emmanuel-Costa,
  %``Geometrical CP Violation,''
  Phys.\ Rev.\ D {\bf 84} (2011) 117901
%  doi:10.1103/PhysRevD.84.117901
  [arXiv:1106.5477 [hep-ph]].

\bibitem{Schwinger:1960qe} 
  J.~S.~Schwinger,
  %``Brownian motion of a quantum oscillator,''
  J.\ Math.\ Phys.\  {\bf 2} (1961) 407.
%  doi:10.1063/1.1703727

\bibitem{Keldysh:1964ud} 
  L.~V.~Keldysh,
  %``Diagram technique for nonequilibrium processes,''
  Zh.\ Eksp.\ Teor.\ Fiz.\  {\bf 47} (1964) 1515
  [Sov.\ Phys.\ JETP {\bf 20} (1965) 1018].

\bibitem{future}
  S.~Pascoli, J.~Turner and Y.-L.~Zhou,
  in progress. 

\bibitem{Calzetta:1986cq} 
  E.~Calzetta and B.~L.~Hu,
  %``Nonequilibrium Quantum Fields: Closed Time Path Effective Action, Wigner Function and Boltzmann equation,''
  Phys.\ Rev.\ D {\bf 37} (1988) 2878.
%  doi:10.1103/PhysRevD.37.2878

\bibitem{Chou:1984es} 
  K.~c.~Chou, Z.~b.~Su, B.~l.~Hao and L.~Yu,
  %``Equilibrium and Nonequilibrium Formalisms Made Unified,''
  Phys.\ Rept.\  {\bf 118} (1985) 1.
%  doi:10.1016/0370-1573(85)90136-X

\bibitem{Berges:2004yj} 
  J.~Berges,
  %``Introduction to nonequilibrium quantum field theory,''
  AIP Conf.\ Proc.\  {\bf 739} (2005) 3
%  doi:10.1063/1.1843591
  [hep-ph/0409233].

\bibitem{Anisimov:2010dk} 
  A.~Anisimov, W.~Buchm\"uller, M.~Drewes and S.~Mendizabal,
  %``Quantum Leptogenesis I,''
  Annals Phys.\  {\bf 326}, 1998 (2011)
  Erratum: [Annals Phys.\  {\bf 338} (2011) 376]
%  doi:10.1016/j.aop.2011.02.002, 10.1016/j.aop.2013.05.00
  [arXiv:1012.5821 [hep-ph]].

\bibitem{Beneke:2010wd} 
  M.~Beneke, B.~Garbrecht, M.~Herranen and P.~Schwaller,
  %``Finite Number Density Corrections to Leptogenesis,''
  Nucl.\ Phys.\ B {\bf 838} (2010) 1
%  doi:10.1016/j.nuclphysb.2010.05.003
  [arXiv:1002.1326 [hep-ph]].

\bibitem{Beneke:2010dz} 
  M.~Beneke, B.~Garbrecht, C.~Fidler, M.~Herranen and P.~Schwaller,
  %``Flavoured Leptogenesis in the CTP Formalism,''
  Nucl.\ Phys.\ B {\bf 843} (2011) 177
%  doi:10.1016/j.nuclphysb.2010.10.001
  [arXiv:1007.4783 [hep-ph]].
  
\bibitem{Prokopec:2003pj} 
  T.~Prokopec, M.~G.~Schmidt and S.~Weinstock,
  %``Transport equations for chiral fermions to order h bar and electroweak baryogenesis. Part 1,''
  Annals Phys.\  {\bf 314} (2004) 208
%  doi:10.1016/j.aop.2004.06.002
  [hep-ph/0312110].

\bibitem{Prokopec:2004ic} 
  T.~Prokopec, M.~G.~Schmidt and S.~Weinstock,
  %``Transport equations for chiral fermions to order h-bar and electroweak baryogenesis. Part II,''
  Annals Phys.\  {\bf 314} (2004) 267
%  doi:10.1016/j.aop.2004.06.001
  [hep-ph/0406140].

\bibitem{Garbrecht:2008cb} 
  B.~Garbrecht and T.~Konstandin,
  %``Separation of Equilibration Time-Scales in the Gradient Expansion,''
  Phys.\ Rev.\ D {\bf 79} (2009) 085003
%  doi:10.1103/PhysRevD.79.085003
  [arXiv:0810.4016 [hep-ph]].

\bibitem{Moore:1995ua} 
  G.~D.~Moore and T.~Prokopec,
  %``Bubble wall velocity in a first order electroweak phase transition,''
  Phys.\ Rev.\ Lett.\  {\bf 75} (1995) 777
%  doi:10.1103/PhysRevLett.75.777
  [hep-ph/9503296].

\bibitem{Moore:1995si} 
  G.~D.~Moore and T.~Prokopec,
  %``How fast can the wall move? A Study of the electroweak phase transition dynamics,''
  Phys.\ Rev.\ D {\bf 52} (1995) 7182
%  doi:10.1103/PhysRevD.52.7182
  [hep-ph/9506475].

\bibitem{Cline:2011mm} 
  J.~M.~Cline, K.~Kainulainen and M.~Trott,
  %``Electroweak Baryogenesis in Two Higgs Doublet Models and B meson anomalies,''
  JHEP {\bf 1111} (2011) 089
%  doi:10.1007/JHEP11(2011)089
  [arXiv:1107.3559 [hep-ph]].
  
\bibitem{Konstandin:2014zta} 
  T.~Konstandin, G.~Nardini and I.~Rues,
  %``From Boltzmann equations to steady wall velocities,''
  JCAP {\bf 1409} (2014) 028
%  doi:10.1088/1475-7516/2014/09/028
  [arXiv:1407.3132 [hep-ph]].

\bibitem{Kozaczuk:2015owa} 
  J.~Kozaczuk,
  %``Bubble Expansion and the Viability of Singlet-Driven Electroweak Baryogenesis,''
  JHEP {\bf 1510} (2015) 135
%  doi:10.1007/JHEP10(2015)135
  [arXiv:1506.04741 [hep-ph]].

%\cite{GarciaBellido:1999sv}
\bibitem{GarciaBellido:1999sv} 
  J.~Garcia-Bellido, D.~Y.~Grigoriev, A.~Kusenko and M.~E.~Shaposhnikov,
  %``Nonequilibrium electroweak baryogenesis from preheating after inflation,''
  Phys.\ Rev.\ D {\bf 60}, 123504 (1999)
  doi:10.1103/PhysRevD.60.123504
  [hep-ph/9902449].
  %%CITATION = doi:10.1103/PhysRevD.60.123504;%%
  %199 citations counted in INSPIRE as of 05 Feb 2018
  
  %\cite{Krauss:1999ng}
\bibitem{Krauss:1999ng} 
  L.~M.~Krauss and M.~Trodden,
  %``Baryogenesis below the electroweak scale,''
  Phys.\ Rev.\ Lett.\  {\bf 83}, 1502 (1999)
  doi:10.1103/PhysRevLett.83.1502
  [hep-ph/9902420].
  %%CITATION = doi:10.1103/PhysRevLett.83.1502;%%
  %113 citations counted in INSPIRE as of 05 Feb 2018
  
  
\bibitem{bellac}
 M. Le Bellac, {\it Thermal Field Theory}, Cambridge University Press 1996. 

\bibitem{PDG}
%\bibitem{Agashe:2014kda} 
  K.~A.~Olive {\it et al.} [Particle Data Group Collaboration],
  %``Review of Particle Physics,''
  Chin.\ Phys.\ C {\bf 38} (2014) 090001.
%  doi:10.1088/1674-1137/38/9/090001

\bibitem{Pilaftsis:2008qt} 
  A.~Pilaftsis,
  %``Electroweak Resonant Leptogenesis in the Singlet Majoron Model,''
  Phys.\ Rev.\ D {\bf 78} (2008) 013008
  %doi:10.1103/PhysRevD.78.013008
  [arXiv:0805.1677 [hep-ph]].
  
\bibitem{Lee:2004we} 
  C.~Lee, V.~Cirigliano and M.~J.~Ramsey-Musolf,
  %``Resonant relaxation in electroweak baryogenesis,''
  Phys.\ Rev.\ D {\bf 71} (2005) 075010
  %doi:10.1103/PhysRevD.71.075010
  [hep-ph/0412354].
  
\end{thebibliography}
\end{document}